\documentclass[aps,prl,10pt,final,twocolumn,showpacs]{revtex4}
\usepackage{graphicx}
\usepackage{array}
\usepackage{amsmath}
\usepackage{amssymb}
\usepackage{upgreek}
\usepackage{units}
\usepackage{color}
\usepackage{float}

\makeatletter
\def\@dotsep{4.5}
\makeatother

\newlength{\colwidth}
\begin{document}

\title{Droplets move over viscoelastic substrates by surfing a ridge}
\date{\today}

\author{S. Karpitschka$^1$}
\email{s.a.karpitschka@utwente.nl}
\author{S. Das$^2$}
\author{M. van Gorcum$^1$}
\author{H. Perrin$^3$}
\author{B. Andreotti$^3$}
\email{andreotti@pmmh.espci.fr}
\author{J.H. Snoeijer$^{1,4}$}
\email{j.h.snoeijer@utwente.nl}
\affiliation{$^1$Physics of Fluids Group, Faculty of Science and Technology, Mesa+ Institute, University of Twente,
7500 AE Enschede, The Netherlands.\\
$^2$Department of Mechanical Engineering, University of Maryland, College Park, MD 20742, USA \\
$^3$Physique et M\'ecanique des Milieux H\'et\'erog\`enes, UMR
7636 ESPCI -CNRS, Univ. Paris-Diderot, 10 rue Vauquelin, 75005, Paris, France\\
$^4$Department of Applied Physics, Eindhoven University of Technology, P.O. Box 513, 5600MB Eindhoven, The Netherlands.
}

\begin{abstract}
Liquid drops on soft solids generate strong deformations below the contact line, resulting from a balance of capillary and elastic forces. The movement of these drops may cause strong, potentially singular dissipation in the soft solid. Here we show that a drop on a soft substrate moves by surfing a ridge: the initially flat solid surface is deformed into a sharp ridge whose orientation angle depends on the contact line velocity. We measure this angle for water on a silicone gel and develop a theory based on the substrate rheology. We quantitatively recover the dynamic contact angle and provide a mechanism for stick-slip motion when a drop is forced strongly: the contact line depins and slides down the wetting ridge, forming a new one after a transient. We anticipate that our theory will have implications in problems such as self-organization of cell tissues or the design of capillarity-based microrheometers.
\end{abstract}

\maketitle

Capillary interactions of soft materials are ubiquitous to nature and play a major role in the self-organization of cell tissues~\cite{ManningPNAS12}, e.g. in embryotic development~\cite{TrinkhausPNAS55,SteinbergSci63}, wound healing~\cite{ArmstrongJCellSci92}, or spreading of cancer cells~\cite{BrochardSM12,SabariPLoSOne11}.
Not least motivated by this, ``soft wetting''~\cite{Shanahan87,PBBB08, JXWD11} recently came to the attention of both, physicists and biologists. Despite its potential for applications, e.g. in the patterning of cells~\cite{Science2005} or droplets~\cite{durotaxisPNAS13} onto soft surfaces, or the optimization of condensation processes~\cite{SARLBB10}, our fundamental understanding especially of the dynamics of soft wetting lags behind by far of what is known about rigid surfaces~\cite{BEIMR09}.

\begin{figure}[t!]
	\centering\includegraphics{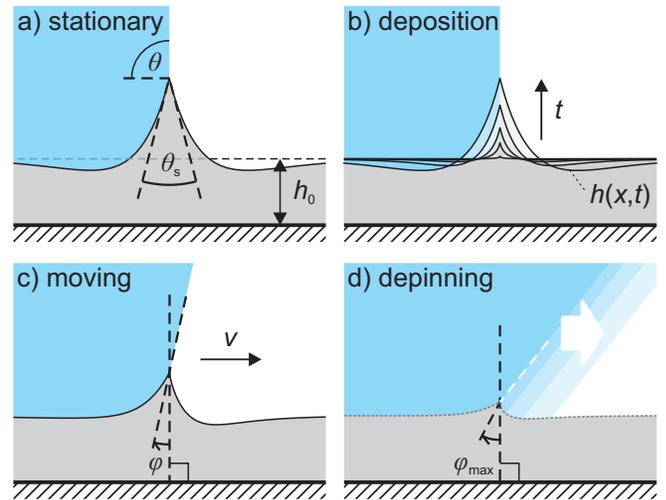}
	\vspace{-5 mm}
	\caption{\label{fig:sketch}\textbf{Dynamics of wetting ridges.} (a) Equilibrium deformation by a three-phase contact line, inducing a solid contact angle $\theta_{\rm s}$. The liquid contact angle is denoted $\theta$. (b) Growth of the ``wetting ridge'' after a drop is deposited. (c) Contact line moving at a velocity $V$. The motion induces a rotation $\varphi$ of the wetting ridge and the liquid contact angle, while $\theta_{\rm s}$ remains constant. (d) Dynamical depinning occurs at a critical angle $\varphi_{\rm depin}.$}
	\vspace{-5 mm}
\end{figure}

Partial wetting of a liquid on a rigid (smooth) substrate is controlled by intermolecular interactions, whose strength is characterized by surface energies~\cite{BEIMR09}. The motion of the three-phase contact line is governed by the viscous dissipation in the liquid. A dissipation singularity arises at the moving contact line~\cite{HS71} and its regularization an the nanoscopic scale can result from various processes~\cite{BEIMR09,SA13}. When the substrate is deformable, a sharp ridge forms below the contact line at the edge of the droplet~\cite{Shanahan87,PBBB08, JXWD11}. The ridge geometry (Fig.~\ref{fig:sketch}a) originates from the coupling between elasticity and surface energy~\cite{Shuttleworth50, LAL96,Limat12,WAS13,BostwickSM14}. The problem is inherently multi-scale and non-local, even at equilibrium, due to the long-range of elastic interactions~\cite{White03,Style12,LUUKJFM14}.

Pioneering experiments  have shown that the softness drastically slows down the wetting dynamics~\cite{SCL95,CGS96} as compared to rigid solids. This ``viscous braking'' has been attributed to a viscoelastic force, as  discussed in several recent experimental articles \cite{Limat13,durotaxisPNAS13}. The theoretical description of moving contact lines over soft solids is so far limited to global dissipation arguments~\cite{LAL96}, which, at least for wetting of rigid solids, are known not to capture the entire physics behind the process.

In this Communication, the physical mechanism that governs soft wetting dynamics is revealed. We measure the dynamical wetting of small water droplets on a rheologically characterized silicone gel and discover a saturation of the dynamical contact angle for large speeds, associated with a maximum friction force. Driving the contact line motion beyond this maximal force eventually leads to a dynamical depinning where the contact line surfs down the wetting ridge, providing a mechanism for recently observed stick-slip motion~\cite{Limat13}. We develop a theoretical framework for dynamical soft wetting, suitable for any substrate rheology. The dynamic wetting angle is calculated from the velocity dependent shape of the deformed solid (Fig.~\ref{fig:sketch}c). The experimental results are matched quantitatively, including saturation/dynamical depinning. The latter arises from an upper limit of the viscoelastic braking effect, which, by exploring different rheologies, is found to be a robust, universal feature of soft wetting and should thus be relevant far beyond droplets on silicone gels. In addition, the analysis captures recent x-ray measurements on the slow growth of wetting ridges when a drop is deposited on a substrate~\cite{PARKNATURE} (Fig.~\ref{fig:sketch}b). 

\section{Results}
\subsection{Experiments.}%
\begin{figure}
\includegraphics{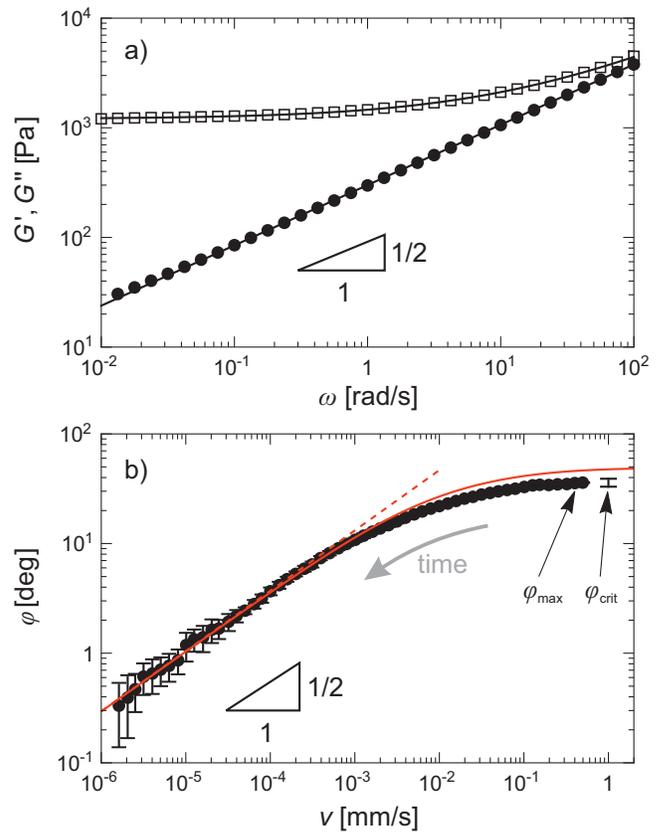}
\vspace{-5 mm}
\caption{\label{fig:thetadyn} \textbf{Rheology of the substrate and dynamic contact angle.} (a) Storage modulus $G'(\omega)$ (open symbols) and loss modulus $G''(\omega)$ (closed symbols) of the silicone gel. Lines are best fit of $\mu\!\!=\!\! G[1+(i \tau \omega)^n]$, giving $n\!\!=\!\! 0.55$, $G \!\!=\!\! \unit[1.2]{kPa}$, and $\tau \!\!=\!\!  \unit[0.13]{s}$. 
(b) Dynamic angle $\varphi = \theta-\theta_{eq}$ for water on the silicone gel (symbols). Data are averaged over 10 independent experiments, error bars represent the standard deviation. The small $v$ behavior exhibits the same power-law as $G''$. Dashed line is the best fit of the asymptote (\ref{eq:thetad}). Solid line corresponds to (\ref{eq:PointWave}), describing the full range of velocities. The diamond indicates $\varphi_{\rm depin}=39^\circ \pm 3^\circ$, which is the separately measured critical angle of depinning.}
\vspace{-5 mm}
\end{figure}

Experiments were performed using water drops on a silicone gel (cf. methods section for details). This system was previously used in static~\cite{Style13} and transient~\cite{PARKNATURE} experiments. Fig.~\ref{fig:thetadyn}(a) shows the rheology of this gel, similar data were reported in \cite{StyleSM14}. The storage $G'$ and loss $G''$ moduli are related by Kramers-Kronig relation: they originate from the same relaxation function $\Psi(t)$. More precisely, the complex shear modulus obey the relation $\mu(\omega)\equiv G' + iG''= i\omega\int_0^{\infty}dt\, \Psi(t)\exp{-i\omega t}$. A silicone gel is a reticulated polymer formed by polymerizing small multifunctional prepolymers: contrarily to other types of gels, there is no liquid phase trapped inside. Such cross-linked polymer networks exhibit scale-invariance that yields power-law response of the form \cite{ChambonWinter, LAL96,deGennes1996}:
\begin{equation}\label{eq:gel}
\Psi(t) = G \left[ 1 +\Gamma(1-n)^{-1}\left( \frac{\tau}{t}\right)^{n} \right],
\end{equation}
where $G$ is a static shear modulus and $\Gamma$ is the gamma function. The associated complex modulus reads $\mu = G'+iG'' =G[1+(i \tau \omega)^n]$. The value of $n$ is not universal but depends on the stoichiometric ratio $r$ between reticulant and prepolymer, with $n$ typically between $2/3$ and $1/2$ \cite{ScanlanWinter}. The best fit in Fig.~\ref{fig:thetadyn}(a) gives an exponent $n=0.55$. Note that the associated effective viscosity $G''/\omega \sim G (\tau/\omega)^n$ is very large, beyond $10\, {\rm Pa}\cdot {\rm s}$ over the entire frequency domain. This will imply that dissipation mainly occurs in the solid, not in the liquid.

Figure~\ref{fig:thetadyn}(b) shows the dynamical angle $\varphi$ as a function of velocity, both of which are measured while the droplet slowly relaxes over time towards its equilibrium shape (after the injection phase). The resulting $\varphi$ versus $v$ is not sensitive to the history of the relaxation process, and the dynamics can thus be considered ``quasi-steady''. The log-log plot reveals a power-law relation between $\varphi$ and $v$ at small velocities, with an exponent equal to the rheological value of $n=0.55$, within error bars. Such a power-law dependence is similar to previously obtained results~\cite{SCL95,CGS96}. For large velocities, we report a striking saturation of the dynamical contact angle. Neither the small-$v$ power law, nor the saturation, can be explained by dissipation mechanisms in the liquid, and one needs to account for the dissipation within the solid. Long et al.~\cite{LAL96} addressed this using a global dissipation argument, based on the solid rheology, but this fails to capture key features such as the saturation.

When the drop is kept inflating with a large over-pressure, we observe a ``depinning'' of the wetting front, with a sudden increase of its velocity, as the dynamical angle reaches a value $\varphi_{\rm depin}\approx 39^{\circ}\pm 3^{\circ}$. This compares well to the saturation of $37^{\circ}$ observed during relaxation (after injection), as indicated in Fig.~\ref{fig:thetadyn}. When forcing the contact angle beyond this angle, the contact line dynamically depins from the wetting ridge, surfing it, until a new ridge forms. Note that such a depinning, leading to stick-slip motions, had already been reported in \cite{Limat13,Limat14}. Our current measurements show that this is a direct consequence of the saturation of the dynamic contact angle.

\subsection{Theoretical framework. }
A liquid drop deposited on a substrate exerts a capillary traction on the surface~\cite{L61,Rusanov75,R78,Shanahan87,bookDeGennes}. While the resulting elastic deformation has been computed and measured for static situations~\cite{PBBB08, Limat12,LUUKJFM14,BostwickSM14,JXWD11,Style13}, the traction becomes time-dependent in the case of dynamical wetting. Here we consider a single straight contact line, for which the elastic problem is two-dimensional. Our goal is to find the deformation of the solid, $h(x,t)$, resulting from the time-dependent capillary traction, $T(x,t)$. For simplicity, we consider the same surface energy $\gamma_{\rm s}$ for the wet and the dry parts of the substrate, and assume that there is no Shuttleworth effect: $\gamma_{\rm s}$ does not depend on strain  \cite{Shuttleworth50, WAS13}. The resulting traction on the solid is then purely normal, and reads $T(x,t)+\gamma_{\rm s} \partial_{xx} h$, the latter term being the solid Laplace pressure~\cite{MPFPP10,MDSA12, JXWD11, LUUKJFM14}. The theory is rigorous for small slopes $(\partial_x h)^2 \ll 1$, but can be extended in a semi-quantitative way to finite slopes.

The shape of the deformed substrate $h(x,t)$ follows from the normal substrate displacements. Inside a purely elastic material, the displacements adapt instantaneously to changes in the capillary traction; the problem is therefore essentially static. For realistic soft materials, however, the displacements are delayed with respect to the imposed forcing. For small deformations, this is captured by a linear stress-strain rate relation  
\begin{equation}
	\label{eq:stressstrain}
	\sigma_{ij}({\bf x}, t) = \int_{-\infty}^{t}d t'\, \Psi(t-t')\partial_{t'} \epsilon_{ij}({\bf x},t') - p({\bf x},t)\delta_{ij}\text{,}
\end{equation}
where $\Psi$ is the relaxation function previously introduced and $p$ is the pressure. Like~\cite{LAL96}, we apply a Fourier transform with respect to time (noted by~``~$\widehat{\ }$~''):
\begin{equation}
	\label{eq:stressstrainfreq}
	\widehat{\sigma}_{ij}({\bf x},\omega) = \mu(\omega) \widehat{\epsilon}_{ij}({\bf x},\omega) - \widehat{p}({\bf x},\omega)\delta_{ij}\text{,}
\end{equation}
The mathematical problem defined by mechanical equilibrium, $\nabla  \cdot   \widehat{\sigma}=0$, the constitutive relation (\ref{eq:stressstrainfreq}) and the traction at the free surface is identical to the static problem, but features dependences on the frequency $\omega$. The time-dependent traction can therefore be solved analogously to~\cite{JXWD11,Style12, LUUKJFM14}, by an additional spatial Fourier transformation (noted by~``~$\widetilde{\ }$~''):
\begin{equation}
	\label{eq:FourierSolution}
	\widehat{\widetilde{h}}(q,\omega) = \frac{\widehat{\widetilde{\mathcal{G}}}(q,\omega)\widehat{\widetilde{T}}(q,\omega)}{1 + \gamma_{\rm s}\,q^2\, \widehat{\widetilde{\mathcal{G}}}(q,\omega)}\text{,}
\end{equation}
where $q$ is the wavenumber. The Green's function $\widehat{\widetilde{\mathcal{G}}}(q,\omega)$ is the product of the time kernel $\mu(\omega)^{-1}$ by the space kernel $K(q)$. For an incompressible layer of thickness $h_0$~\cite{XU10} it is
\begin{equation}
  \label{eq:green}
K(q)=\left[\frac{\sinh(2qh_0) - 2qh_0}{\cosh(2qh_0) + 2(qh_0)^2+1 }\right]\;\frac{1}{2q}
\end{equation}
Left-right symmetry and volume conservation are reflected by $K(q)=K(-q)$ and $K(0)=0$. Sharp features in the solid profile, like the solid contact angle, are found in the large $q$ asymptotics for which $K(q)\simeq (2|q|)^{-1}$. 

\subsection{The moving contact line. }
We now apply our theory to a contact line moving at a constant velocity $v$, which induces a traction
\begin{equation}
\label{eq:traction}
T(x,t) = \gamma\sin\theta\,\delta(x-vt)\text{.}
\end{equation}
This reflects the normal force per unit contact line that is exerted by the liquid on the solid, while $\theta$ is the liquid angle at the location of the cusp. For simplicity we consider that the drop size is much larger than the substrate thickness, in which case the Laplace pressure inside the liquid can be neglected \cite{Style13}. We indeed verified that the finite drop size has a negligible influence on the resulting motion: the relevant scale for the dynamics is the size of the ridge $\gamma_{\rm s}/G$, which is much smaller than the drop size. This also justifies a two-dimensional model. Another important simplification comes from the quasi-steady nature of the droplet relaxation, e.g. temporal changes of contact angle and contact line velocity are small in our experiments ($d\theta/dt \ll \tau^{-1})$, such that the process can be modeled by a constant velocity.

According to (\ref{eq:FourierSolution}), the capillary traction induces a wetting ridge moving at a velocity $v$ (cf. methods section):
\begin{equation}
\label{eq:PointWave}
\widetilde{h}(q) = \frac{\gamma\sin \theta}{\gamma_{\rm s}} \left[q^2 + \frac{\mu(qv)/\gamma_{\rm s}}{K(q)} ) \right]^{-1}\text{,}
\end{equation}
in the comoving frame. In real-space this gives profiles such as shown in Fig.~\ref{fig:sketch}(c). The motion induces a left-right symmetry breaking: the asymmetric deformation of the solid results in a tilt angle $\varphi(v)$ of the cusp. Since the liquid is close to equilibrium, because the dominant dissipation takes place inside the solid, the change in the solid angle $\varphi$ directly yields a change in the liquid angle $\theta$ (Fig.~\ref{fig:sketch}d). Hence $\theta=\theta_{\rm eq}+\varphi$, where $\theta_{\rm eq}$ is the equilibrium liquid angle by Neumann's law. The liquid contact angle $\theta$ gets deviated away from $\theta_{\rm eq}$ due to the viscoelastic forces in the substrate.

Fig.~\ref{fig:thetadyn}(b) shows the calculated tilt curve for the gel used in experiments. It predicts not only the power-law behavior at low velocity, but also presents a saturation of the tilt angle. The tilt angle quantifies the velocity-dependent viscoelastic force between the solid and liquid phases. For a well-established moving ridge, it behaves as a resistive force increasing with the velocity. When the drop is forced to inflate with a driving force larger than the maximal braking force, the contact line can no longer remain ``pinned'' to the steadily moving solid ridge and surfs the gel wave. To investigate further the relation between the tilt $\varphi$ and the substrate constitutive relation, we use the gel-rheology (\ref{eq:gel}), and expand (\ref{eq:PointWave}) in the small $v$ asymptotics (and hence small $\varphi$ i.e., $\sin\theta \approx \sin\theta_{\rm eq}$), which gives (cf. methods section):
\begin{equation}
\label{eq:thetad}
\varphi = \frac{2^{n-1}n}{\cos(n\pi/2)} \, \frac{\gamma\sin\theta}{\gamma_{\rm s}}\, \left( \frac{v}{v^*}\right)^{n} ,
\end{equation}
where the characteristic velocity scale emerges as $v^*=\gamma_{\rm s}/(G\tau)$. Note that the outer length scale (thickness of substrate) does not appear. This expression can be simply interpreted. At vanishing response time $\tau$, a deformation matching the static ridge would propagate at a velocity $v$, pushing the substrate material up and down at a characteristic frequency $\omega$ equal to the velocity $v$ divided by the characteristic width of the ridge $\sim \gamma_{\rm s}/G$. The perturbation introduced by a finite $\tau$ is encoded by the loss modulus $G''(\omega)$. As the characteristic strain is set by the slope of the ridge $\sim \gamma \sin \theta /\gamma_{\rm s}$, one obtains dimensionally Eq.~(\ref{eq:thetad}). The scaling law $\varphi \propto (v/v^*)^n$ thus simply carries over the low frequency behavior $G''(\omega) \propto \omega^n$, which is a robust mechanism valid beyond the small slope approximation of our theory. At small $v$ (small $\omega$), dissipation will dominantly occur in the solid because $n<1$, while the loss modulus of a newtonian liquid $G''_{liq}\propto\omega$.

In contrast to the tilt angle $\varphi$, we find that that the solid angle $\theta_{\rm s}$ does not depend on $v$. This result can be derived analytically from the large-$q$ asymptotics of (\ref{eq:PointWave}), $\tilde{h} \simeq \frac{\gamma \sin\theta}{\gamma_{\rm s}} q^{-2}$, valid for all $v$ and arbitrary $\mu(\omega)$. In real space, this implies a slope discontinuity
\begin{equation}
\label{eq:ThetaS}
\theta_{\rm s} \simeq \pi - \gamma \sin\theta/\gamma_{\rm s}\text{,}
\end{equation}
which is the small-slope limit ($\gamma/\gamma_{\rm s} \ll 1$) of Neumann's contact angle law. Physically, (\ref{eq:ThetaS}) reflects that $\theta_{\rm s}$ is determined by surface tensions only~\cite{JXWD11,MDSA12b,Limat12}: bulk viscoelastic stresses are not singular enough to contribute to the contact angle selection. This feature remains true for arbitrary angles~\cite{LUUKJFM14}. In order to match Neumann's law quantitatively, our theory must be corrected at large slopes to take geometric nonlinearities into account. This can be achieved phenomenologically in (\ref{eq:FourierSolution}) by replacing $\gamma_{\rm s} \rightarrow \sqrt{\gamma_{\rm s}^2- \gamma^2/4}$. Indeed, the Neumann condition for $\theta=90^\circ$ reads $2\gamma_{\rm s} \sin \alpha =\gamma$, where $\alpha$ is the angle of the solid interface with the horizontal. Small-slope theory gives $2|h'|=2\tan \alpha = \frac{\gamma}{\gamma_{\rm s}}$, and hence lacks a factor $\cos \alpha = \sqrt{1-(\gamma/2\gamma_{\rm s})^2}$.

For the first time, we reveal that the exponent of the dynamical contact angle directly originates from the gel rheology. The rheological parameters being calibrated independently, the dynamic contact angle can be fitted to the model to extract the solid surface tension. Using Eq.~(\ref{eq:thetad}), which is valid for small slopes, we find $\gamma_{\rm s} = \unit[16]{mN/m}$. This is a reasonable value, though a bit lower than the value previously derived from Neumann's law \cite{Style13}. We think this difference can be attributed to the small-slope nature of our theory: condering the phenomenological correction for large slopes gives a value $\gamma_{\rm s} = \unit[39]{mN/m}$, in close agreement with \cite{Style13}. The solid line in Fig.~\ref{fig:thetadyn}(b) shows the prediction from (\ref{eq:PointWave}), providing an excellent description over the full range of velocities. The model captures also the saturation, though the value for $\varphi_{\rm crit}$ is slightly overestimated.

\subsection{Depinning and growth of a new wetting ridge. }
How can the contact line escape pinning, without dragging the capillary wedge along with it? To answer this question, let us consider the recent experiments investigating the growth of a wetting ridge after depositing a droplet on a silicone gel \cite{PARKNATURE}. The substrate was observed to only very slowly establish the final shape of the wetting ridge -- such a delay in growth (or decay) of wetting ridges would explain how a sufficiently rapid contact line could escape from the ridge. However, the solid angle $\theta_{\rm s}$ (cf. Fig.~\ref{fig:sketch}a) appeared very quickly and remained constant during the entire growth of the ridge \cite{PARKNATURE}. 

These features of ridge growth can all be explained by considering our theory for a traction that is suddenly imposed at the time of deposition ($t=0$), so that  
\begin{equation}\label{eq:traction1}
T(x,t) = \gamma \sin \theta\; \delta(x) \, \Theta(t)
\end{equation}
where $\theta$ is the liquid contact angle and $\Theta(t)$ is the Heaviside step function. Combining this traction with (\ref{eq:FourierSolution}), one can compute the resulting $h(x,t)$ for any rheology $\mu(\omega)$. An example of the evolution of the wetting ridge is shown in Fig.~\ref{fig:sketch}(b). A movie is given in the Supplementary material (Supplementary Movie 1). 
\begin{figure}
\centering\includegraphics{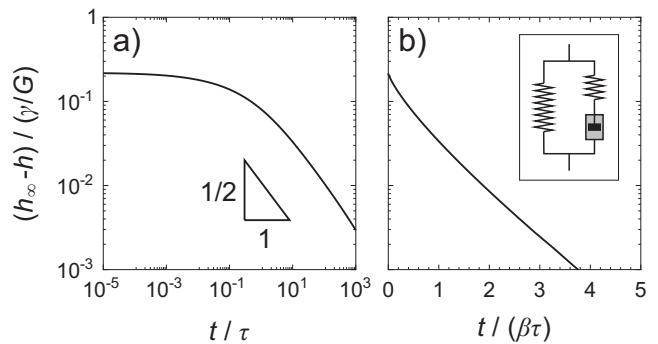}
\vspace{-5 mm}
\caption{\label{fig:onset}\textbf{Relaxation of the central height of the wetting ridge after drop deposition.} The curves show the approach to equilibrium height, $h_{\infty}-h$ at $x=0$, for two rheologies: (a) Power-law rheology ((\ref{eq:gel}) with $n=1/2$) and (b) standard linear model, $\beta = 300$ ((\ref{eq:PsiSTD}), see inset). The dimensionless substrate thickness for these plots was $\gamma_{\rm s}/(G h_0)=0.5$.}
\vspace{-5 mm}
\end{figure}

First, the theory recovers the experimental finding that $\theta_{\rm s}$ is constant at all times. As for the moving contact line, this result can be derived analytically from the large-$q$ asymptotics of (\ref{eq:FourierSolution}), which again results in Eq.~(\ref{eq:ThetaS}): the asymptotics are independent of the rheology and the history of the traction, but entirely governed by the surface tensions.

Second, the theory explains why, contrarily to the rapid appearance of $\theta_{\rm s}$, the global shape of the ridge evolves much more slowly. Figure~\ref{fig:onset} shows the evolution of the central height of the ridge, $h(x=0)$, towards its static value $h_\infty$, for the two idealized rheological models. The relaxation towards the equilibrium height is algebraic for the gel model, with an exponent directly following that of rheological relaxations (Fig.~\ref{fig:onset}(a), as $t^{-1/2}$ for $n=1/2$). This clarifies the complex evolution of the wetting ridge of the silicone gel in \cite{PARKNATURE}: small-scale characteristics like $\theta_{\rm s}$ are dominated by surface tension and relax quickly, while large scale features inherit the relaxation dynamics from the bulk rheology. This means that immediately after depinning, where the contact line exhibits a rapid motion, the solid cusp cannot adapt quickly. The liquid will slide down the wetting ridge, which appears ``frozen'' on the timescale of the depinning. During this phase it is clear that the liquid dynamics will be important -- still, the onset of the depinning can be explained quantitatively without invoking the fluid dynamics inside the liquid, because the saturation angle $\varphi_{\rm max}$ coincides with the observed depinning angle $\varphi_{\rm crit}$ [Fig.~\ref{fig:thetadyn}(b)].

\subsection{Robustness and interpretation. }
The theory can be applied to any viscoelastic substrate, assuming that it is probed in the linear regime. Generic reticulated polymer networks possess a long time entropic elasticity \cite{Flory1943,Rubinstein2002}, that is characterized by a static shear modulus $G$. Such networks become viscoelastic when excited over time-scales shorter than a certain response time $\tau$. In order to investigate the robustness of the phenomenology that was observed experimentally and reproduced quantitatively by our model, we will consider a different rheological limit. When cross-linking long polymer chains, one forms an elastomer rather than a gel. Assuming a single (Rouse) timescale $\tau$ to characterize the onset of entanglements, the rheology can be idealized as~\cite{LAL96,deGennes1996,Rubinstein2002}:
\begin{equation}\label{eq:PsiSTD}
 \Psi(t) = G(1+\beta e^{-t/\tau}).
\end{equation}
This single timescale response is also referred to as standard linear model and has frequently been used to describe the transition from rubber to glass behaviors. In general, several relaxation times must be introduced to capture quantitatively the rheology of actual elastomers.
\begin{figure}
	\centering\includegraphics{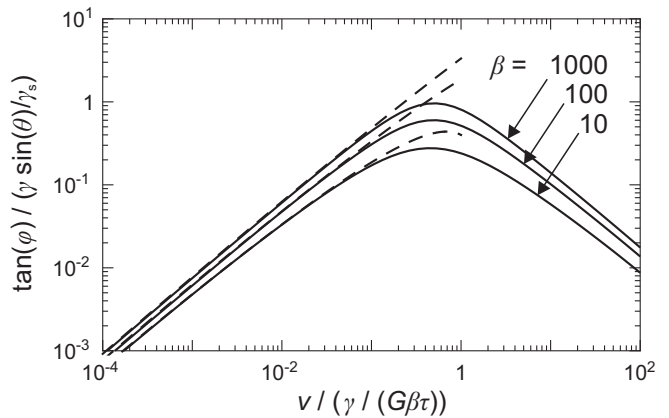}
	\vspace{-5 mm}
	\caption{\label{fig:PointWaveSlopesSTD}\textbf{Cusp tilt for standard linear model.} Solid lines: numerical results; dashed lines: analytical approximation (\ref{eq:thetadKV}). The upper bound for the viscous braking force is robust with respect to the details of the rheology.}
	\vspace{-5 mm}
\end{figure}

The wetting ridge relaxation, which follows the rheological relaxation, becomes exponential for the standard linear model (Fig.~\ref{fig:onset}b). The case of a moving contact line with the rheology (\ref{eq:PsiSTD}) is given by the solid lines in figure~\ref{fig:PointWaveSlopesSTD}, showing the tilt curves for various parameters $\beta$. As for the gel case, the tilt has an upper bound and this therefore appears a robust feature of soft wetting. Note that the maximum depends on $\gamma/\gamma_{\rm s}$ and $\beta$ (or $n$ for the gel case), illustrating that the value of $\varphi_{\rm max}$ will in general depend on the details of the rheology (e.g. $\varphi_{\rm max}$ in \cite{Limat13} is much smaller than in the presented experiments). The dynamic contact angle for a standard linear solid can actually be captured in a simple analytical form. For this we consider the limit $\beta\rightarrow\infty$ while keeping $\tau'\equiv \beta\tau$ constant -- this corresponds to the Kelvin-Voigt model with a frequency-independent effective viscosity $\eta = G\tau'$. Intriguingly, this limit turns out to be singular: the high-frequency behaviour of (\ref{eq:PsiSTD}) becomes purely viscous and gives a non-integrable singularity of the dissipation. This singularity could already be anticipated from (\ref{eq:thetad}), since the Kelvin-Voigt rheology has $G'' \sim \omega^1$, while (\ref{eq:thetad}) presents a divergence for $n=1$.

In fact, this viscoelastic singularity is the soft-solid analogue of the classical Huh \& Scriven-paradox for viscous contact line motion~\cite{HS71}. This is demonstrated from the large-$q$ asymptotics of (\ref{eq:PointWave}) in the Kelvin-Voigt limit, giving a slope close to the contact line (cf. Supplementary~Note~1):
\begin{equation}
	\label{eq:PointWaveKVLargeQSlopes}
		\partial_x h \simeq  - \frac{\gamma\sin\theta}{2\pi\gamma_{\rm s}}\left(
		                       \frac{|x|/x - 4\,\left(\gamma{\rm E}+\ln|x|/h_0\right)\,v/v^*}
		                            {1+4\,(v/v^*)^2}
		                       \right)\text{,}
\end{equation}
with $\gamma{\rm E} =$ Euler's constant and $v^* = \gamma_{\rm s}/(G\tau')$. This expression reveals a logarithmic divergence of the slope, in perfect analogy to the Cox-Voinov result for liquid contact lines~\cite{V76,C86}. 

\begin{figure}
	\centering\includegraphics{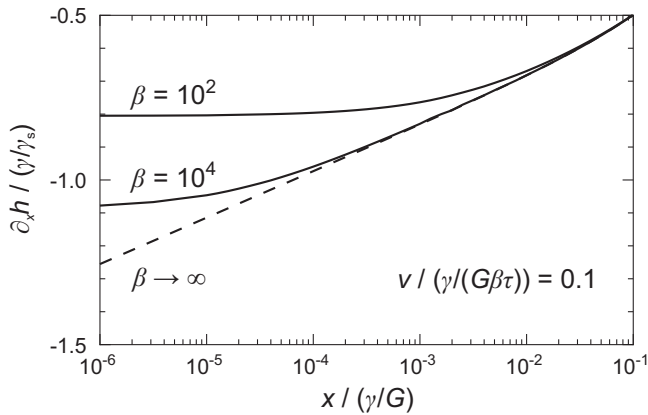}
	\vspace{-5 mm}
	\caption{\label{fig:PointWaveSlopeReg}\textbf{Logarithmic variation of the profile slope.} $\partial_x h$ is plotted as a function of the distance to the contact line. Different curves correspond to different instantaneous relaxation moduli $\beta E$, for identical dimensionless velocity $v=1$. In the limit $\beta \rightarrow\infty$ (the Kelvin-Voigt solid), the slope diverges logarithmically at the contact line. For finite $\beta$, the slope saturates at a regularization length given by (\ref{eq:lreg}).}
	\vspace{-5 mm}
\end{figure}

Contrarily to the viscous-liquid singularity, the presence of an instantaneous elastic response, i.e. a finite value of $\beta$, is sufficient to regularize the divergence. This is illustrated in Fig.~\ref{fig:PointWaveSlopeReg}, which shows the slope ahead of the moving contact line as a function of the distance to the contact line. The dashed line corresponds to the Kelvin-Voigt limit (\ref{eq:PointWaveKVLargeQSlopes}), showing the logarithmic steepening of the slope. For finite $\beta$, the slope saturates upon approaching the moving contact line. The saturation wavenumber is found $q' \sim (v\tau)^{-1}$, which corresponds to a length
\begin{equation}
\label{eq:lreg}
\ell = \tau v\text{.}
\end{equation}
$\ell$ is a dynamical regularization length that depends linearly on the velocity of the contact line; for $v \sim {\cal O}(1)$ this scale is still much smaller than the substrate thickness, by a factor $\beta^{-1}$. The physical origin of the regularization lies in the instantaneous elasticity in the high-frequency limit, which applies at frequencies beyond $ \sim \beta \tau$.

Inserting the regularization length into the Kelvin-Voigt limit (\ref{eq:PointWaveKVLargeQSlopes}), we identify the tilt and get an analytical expression of the dynamic liquid contact angle (the strict validity of the analysis requires small slopes, i.e. small $\varphi$; we therefore replaced $\sin \theta = \sin(\pi/2 + \varphi) \simeq 1$):
\begin{equation}\label{eq:thetadKV}
	\theta = \theta_{\rm eq} + \,\frac{2}{\pi} {\rm Ca}_{\rm s} \ln\left(\frac{h_0}{e^{\gamma{\rm E}}\ell}\right) \text{.}
\end{equation}
This result is analogous to the Cox-Voinov law~\cite{V76,C86} in partial wetting of viscous fluids. In that case a similar logarithmic factor linking microscopic and macroscopic scales appears, for arbitrary contact angles \cite{C86}, and the resulting expression for small Ca is of the form~(\ref{eq:thetadKV}).
Interestingly, the analysis reveals that the relevant dimensionless velocity for soft wetting is not the classical liquid capillary number ${\rm Ca}=v\eta_\ell/\gamma$, based on the liquid viscosity $\eta_\ell$, but the ``solid capillary number''
\begin{equation}
{\rm Ca}_{\rm s} = \frac{v G\beta\tau \gamma }{\gamma_{\rm s}^2}\text{.}
\end{equation}
Equation (\ref{eq:thetadKV}) closely follows the numerical results (Fig.~\ref{fig:PointWaveSlopesSTD}(d), dashed and solid lines, respectively). The moving contact line singularity is avoided altogether when $G''$ has an exponent $n<1$, as was the case for the power-law gel, in perfect analogy to shear-thinning fluids moving on a rigid substrate.

\section{Discussion} 
We have shown how contact lines can surf on a wetting ridge, and that this governs the remarkable spreading of drops on viscoelastic substrates. We have quantified this dynamics by measuring the dynamic contact angle of water on a PDMS gel for a wide range of velocities and described, for the first time, a saturation of the dynamical angle for large velocities. This saturation is in harsh contrast to wetting dynamics of rigid solids and leads to depinning, where the contact line slides down the ridge until a new wetting ridge has had time to grow and sustain a steady motion -- hence, explaining the remarkable stick-slip motion~\cite{Limat13} found recently on soft solids. We develop a theory that identifies a robust maximum in viscous braking force that correctly predicts the onset of dynamical depinning. In addition, our theory captures the unsteady growth of a wetting ridge \cite{PARKNATURE}.
This work provides a framework for viscoelasto-capillary dynamics valid beyond droplets, and should be applicable e.g. within a biological context. It also opens a new perspective where droplets can be used droplets as Microrheometers, since I) the length scales probed by the droplets is given by the elastocapillary length i.e., a few microns, and II) the tilt saturation occurs at velocities that are directly related to the relaxation timescale.

\section{Methods} 
\subsection{Wetting experiments.} The silicone gels (Dow Corning CY52-276) are prepared by curing the mixed components onto glass slides, yielding $0.8$ mm thick substrates. The rheology was determined using a MCR 501 rheometer (Anton Paar). Dynamic contact angles were measured using droplets of MilliQ water dispensed from a clean Hamilton syringe. First, a small droplet ($\approx\unit[2...20]{\upmu l}$) was placed onto the substrate, leaving the syringe needle attached to the droplet. Then, the contact angle of the droplet was increased by quickly injecting water ($\approx\unit[3...20]{\upmu l}$ with $\approx\unit[2...8]{\upmu l s^-1}$) to it. After the injection phase the drop relaxes quasi-statically, causing the contact line velocity to decay slowly. The advancing motion of the contact line and the relaxation of the contact angle were imaged at $50$~Hz with a long distance video microscope. The droplet contour was extracted with sub-pixel resolution, and velocities down to $\unit{nm/s}$ could be detected. The measured contact angles were translated to tilt angles $\varphi$ by subtracting $\theta_{\rm eq} \approx106^{\circ}\pm1^{\circ}$.

\subsection{The moving contact line}
Fourier transforming (\ref{eq:traction}) from $x$ to $q$ and from $t$ to $\omega$ preserves the $\delta$-shape of the traction:
\begin{align}
	\widehat{\widetilde{T}}(q,\omega) = 2\pi\,\gamma\,\sin\theta\,\delta(\omega - vq)\text{.}
	\label{eq:Thattilde}
\end{align}
Inserting the above into Eq.~(\ref{eq:FourierSolution}), the inverse transform to the time domain yields
\begin{align}
	\widetilde{h}(q,t) = \frac{\gamma\sin\theta}{\gamma_{\rm s}}\left[ q^2 + \frac{\mu(vq)}{K(q)} \right]^{-1}\,e^{iqvt}\text{.}
	\label{eq:htilde}
\end{align}
The only explicit time dependence appears in the phase factor that shifts the profile in $x$-direction linearly with time. The transformation to the co-moving frame is done by multiplication with $e^{-iqvt}$, which cancels the only explicit time dependence, and one obtains Eq.~(\ref{eq:PointWave}).

The slopes are evaluated by multiplication with $-iq$ prior to the inverse transform (in the co-moving frame):
\begin{align}
	h'(x) = \frac{1}{2\pi}\int_{-\infty}^{\infty}\left( -iq \widetilde{h}(q)\right)\,e^{-iqx}\,dq\text{.}
	\label{eq:hprime}
\end{align}
$h'(x)$ is a real function because $\Re\left[-iq\widetilde{h}(q)\right] = \Re\left[iq\widetilde{h}(-q)\right]$ and $\Im\left[-iq\widetilde{h}(q)\right] = -\Im\left[iq\widetilde{h}(-q)\right]$.
$h'(x)$ can be split into a symmetric and an antisymmetric part, where the symmetric part is given by the inverse transform of the real part of $iq\widetilde{h}(q)$:
\begin{equation}
	\frac{1}{2}\left(h'(x) + h'(-x)\right) = \frac{1}{2\pi}\int_{-\infty}^{\infty}\Re\left[-iq\widetilde{h}(q)\right]\,e^{-iqx}\,dq\text{,}
\end{equation}
The antisymmetric part is obtained form the imaginary part:
\begin{equation}
	\frac{1}{2}\left(h'(x) - h'(-x)\right) = \frac{1}{2\pi}\int_{-\infty}^{\infty}\Im\left[-iq\widetilde{h}(q)\right]\,e^{-iqx}\,dq\text{.}
\end{equation}

The solid angle $\theta_{\rm s}$ is given by the (antisymmetric) slope discontinuity at $x=0$ and is thus encoded in the backward transform of the imaginary part. The discontinuity is caused by the large-$q$ asymptotics alone. If $\mathcal{O}(\mu(v\,q))<\mathcal{O}(q)$, which is the case for the exponential- and power law ($n<1$) relaxation (but not for the Kelvin-Voigt model), it is independent of rheology:
\begin{align}
	&\lim_{x\rightarrow 0^+}h'(x) - \lim_{x\rightarrow 0^-}h'(x) \nonumber\\
	&= \frac{1}{\pi}\left(\lim_{x\rightarrow 0^+}\int_{-\infty}^{\infty}\frac{-i \gamma \sin\theta}{\gamma_{\rm s} q}\,e^{-iqx}\,dq\right) = -\frac{\gamma \sin\theta}{\gamma_{\rm s}}\text{.}
\end{align}

The rotation of the wetting ridge is given by the symmetric part of $h'(x)$ and thus obtained by the backward transform of the real part, evaluated at $x=0$:
\begin{align}
	\tan\varphi &= \lim_{x\rightarrow0}\frac{1}{2}\left(h'(x) + h'(-x)\right)\nonumber\\
		&=\frac{1}{2\pi}\int_{-\infty}^{\infty}\Re\left[-iq\widetilde{h}(q)\right]dq
	\label{eq:tanphi}
\end{align}
With the symmetry property $K(q)=K(-q)$, and small, positive $v$, Eq.~(\ref{eq:tanphi}) simplifies to (primes omitted):
\begin{equation}
	\varphi = \frac{\gamma\sin\theta}{G}\,\frac{\sin\nicefrac{n\pi}{2}}{\pi}\int_0^{\infty}
						\frac{ q(qv\tau)^{n}K(q)}
								 { \left((\gamma_{\rm s}/G) q^2 K(q) + 1\right)^2}
		\,dq\text{.}
\end{equation}
In the limit of thick elastic layers, $K(q) = (2|q|)^{-1}$. After non-dimensionalizing the integration variable as $q'=\frac{\gamma_{\rm s}}{G} q$, one obtains (primes omitted):
\begin{align}
	\varphi &= \frac{\gamma\sin\theta}{\gamma_{\rm s}}\left(\frac{v}{\gamma_{\rm s}/(G\tau)}\right)^n \frac{\sin\nicefrac{n\pi}{2}}{2\pi}\int_0^{\infty}
		\frac{ q^n}
				 { \left(q/2 + 1\right)^2}
		\,dq\nonumber\\
		&=\frac{2^{n-1} n}{\cos\nicefrac{n\pi}{2}}\,\frac{\gamma\sin\theta}{\gamma_{\rm s}}\left(\frac{v}{v^*}\right)^n\text{,}
\end{align}
where $v^* = \gamma_{\rm s}/(G\tau)$ is the characteristic velocity.

\subsection{Growth of a wetting ridge.}
Here we give the full derivation of the time-dependent wetting ridge shape after the deposition of a droplet. We only discuss the result for the exponential relaxation model. An analogous calculation can be performed for the power-law relaxation.

In the following, we non-dimensionalize $x$ with $h_0$, $q$ with $h_0^{-1}$, $t$ with $\beta\tau$, $\omega$ with $(\beta\tau)^{-1}$, and $h$ with $\gamma \sin\theta\, /G$.
With this scaling, the Fourier transform of the time-kernel for exponential relaxation (\ref{eq:PsiSTD}) reads
\begin{equation}
	\mu(\omega) = G\left(1+\frac{\omega}{\omega/\beta-i}\right)\text{.}
\end{equation}
The space kernel in scaled variables is
\begin{equation}
	\kappa(q) = \frac{K(q)}{h_0} = \left[\frac{\sinh(2q) - 2q}{\cosh(2q) + 2q^2+1 }\right]\;\frac{1}{2q}\text{.}
\end{equation}
The traction Eq.~(\ref{eq:traction1}) is transformed to
\begin{equation}
	\widehat{\widetilde{T}}(q,\omega)=\frac{\gamma\sin\theta}{h_0}\left(\frac{i}{\omega}+\pi\delta(\omega)\right)\text{.}
\end{equation}
Equations (\ref{eq:Thattilde}), (\ref{eq:htilde}), and (\ref{eq:hprime}) are inserted into the general expression Eq.~(\ref{eq:stressstrainfreq}), which yields:
\begin{equation}
	  \widehat{\widetilde{h}}(q,\omega)
	= \frac{\left( \frac{i}{\omega}+\pi\delta(\omega) \right)}
				 {\left(1+\frac{\omega}{\omega/\beta-i}\right){\kappa(q)}^{-1} + \alpha_{\rm s} q^2}\text{,}
\end{equation}
with the dimensionless parameter $\alpha_{\rm s} = \gamma_{\rm s}/(G h_0)$. $\alpha_{\rm s}$ compares the elastocapillary length for the solid surface tension to the layer thickness $h_0$.
The inverse Fourier transform to the time-domain yields
\begin{equation}
	  \widetilde{h}(q,t)
	= \frac{1 - \frac{\beta \exp\left[ - \frac{1+\alpha_{\rm s} q^2 \kappa(q)}{1+\beta+\alpha_{\rm s} q^2 \kappa(q)} \beta t \right]}
													 {1+\beta+\alpha_{\rm s} q^2 \kappa(q)}}
	       {{\kappa(q)}^{-1} + \alpha_{\rm s} q^2}\text{.}
\end{equation}

Fourier transformation to real space is performed numerically.

%
%
%

\bibliographystyle{naturemag}

%
%
\section{Acknowledgements}
We thank T. Baumberger, A. Eddi and E. Raphael for fruitful discussions. SK acknowledges financial support from NWO through VIDI Grant No. 11304. JS acknowledges financial support from ERC (the European Research Council) Consolidator Grant No. 616918.

\section{Author contributions}
JS and BA designed the research, SK, JS, and SD developed the theory, MG and SK performed and analyzed the wetting experiments, HP performed the rheological measurements, BA and JS wrote the Manuscript. All authors discussed the results and commented for the final manuscript.

\section{Additional information}
\subsection{Supplementary information}
Supplementary Movie 1: The velocity dependent shape of the wetting ridge.
\subsection{Competing financial interests}
The authors declare no competing financial interests.
%
%
%
%
\end{document}